\documentclass{aastex701}
\usepackage{CJK}
\usepackage{listings}

\newcommand{\HI}{H\,{\sc i}}
\newcommand{\Rv}{R_{\rm V}}
\newcommand{\Tsnr}{\theta_{\rm SNR}}
\newcommand{\Msun}{M_\odot}
\newcommand{\Rsnr}{R_{\rm SNR}}
\newcommand{\tsnr}{t_{\rm SNR}}
\newcommand{\Rsun}{R_\odot}
\newcommand{\dRv}{\Delta R_{\rm V}}
\newcommand{\rs}{r_{\rm s}}

\newcommand{\MswDust}{M_{\rm swept}^{\rm dust}}
\newcommand{\MejDust}{M_{\rm ej}^{\rm dust}}
\newcommand{\avea}{\langle a \rangle}

\submitjournal{ApJL}
\begin{document}
\begin{CJK*}{UTF8}{gbsn}

\title{Observational Evidence of Dust Evolution in Supernova Remnants: Size Redistribution Toward Larger Grains in Early Sedov Phase}

\author[0000-0003-2645-6869]{He Zhao (赵赫)}
\affiliation{Departamento de Fisica y Astronomia, Facultad de Ciencias Exactas, Universidad Andres Bello, 
Fernandez Concha 700, 8320000 Santiago, Chile}
\email[show]{he.zhao@oca.eu}

\author[0000-0003-2472-4903]{Bingqiu Chen (陈丙秋)}
\affiliation{South-Western Institute for Astronomy Research, Yunnan University, Kunming, Yunnan 650091, 
People's Republic of China}
\email[show]{bchen@ynu.edu.cn}

\author[0000-0001-9328-4302]{Jun Li (李军)}
\affiliation{Center for Astrophysics, Guangzhou University, Guangzhou 510006, People's Republic of China}
\email[show]{lijun@gzhu.edu.cn}

\begin{abstract}

Recent observations have revealed that dust is widespread and abundant in galaxies up to $z\,{\backsimeq}\,8$, significantly 
influencing their appearance and spectral properties. In the early Universe, dust is thought to form primarily in 
supernova (SN) ejecta, but also undergoes destruction by the reverse shock. Studying dust in local supernova remnants 
(SNRs) of different sizes and ages thus provides key constraints on dust formation and evolution during cosmic dawn.
Using the newly released 3D $\Rv$ map, we derived local $\Rv \equiv A_{\rm V}/E(B-V)$ values for dust in 14 Galactic 
SNRs in the early Sedov phase and their surrounding interstellar medium (ISM). For the first time, we detect a moderately 
strong positive correlation between the difference in SNR and ISM $\Rv$ ($\dRv$) and SNR radius ($\Rsnr$), with a 
Spearman coefficient of $\rs\,{=}\,0.62\,{\pm}\,0.14$. This trend offers direct observational evidence for a 
redistribution of dust grain sizes toward larger grains during SN shock processing, consistent with theoretical models. 
Our findings provide essential observational constraints on dust size evolution in SNRs and important implications for 
understanding the rapid enrichment and survival of dust in the early Universe.
% ---------------------------------------------------------------------------------------------------------------------

\end{abstract}


\section{Introduction} \label{sect:introduction}

Dust is a fundamental component of the interstellar medium (ISM), profoundly shaping the spectral energy distribution 
(SED) of galaxies. By attenuating ultraviolet, optical, and near-infrared (IR) radiation and re-emitting the absorbed 
energy at mid-/far-IR and millimeter wavelengths, dust not only modifies galaxy spectra but also plays a crucial role 
in interstellar chemistry and star formation processes (see reviews by \citealt{Draine2003}, \citealt{HD2021}, and 
\citealt{Galliano2018}). Remarkably, substantial quantities of dust have been discovered in very distant galaxies at 
redshifts up to $z\,{\backsimeq}\,8$ \citep[e.g.][]{Laporte2017, Witstok2023}, corresponding to the first billion years 
after the Big Bang. These observations significantly reshape our understanding of cosmic evolution. Consequently, 
determining how dust originated and rapidly evolved during the earliest phases of galaxy formation has become a central 
focus in astrophysics (see \citealt{SM2024} for a review).
% ---------------------------------------------------------------------------------------------------------------------

Studying dust evolution in nearby Galactic supernova remnants (SNRs) provides crucial insights into dust formation during 
cosmic dawn. Supernovae (SNe) are widely considered the primary dust producers in the early Universe \citep{Sarangi2018}, 
with substantial dust detected in SN ejecta across both Galactic and extragalactic SNRs. Notable examples include the Crab 
Nebula \citep{Gomez2012, TD2013, OB2015}, Cassiopeia A \citep[Cas A;][]{Dwek1987, Barlow2010, DeLooze2017}, and SN 1987A 
\citep{Matsuura2011, Indebetouw2014, Matsuura2015, Wesson2015, BB2016}. However, dust observed in young SNRs does not 
represent the final quantity injected into the ISM, as significant processing occurs when ejecta encounter reverse shocks.
% ---------------------------------------------------------------------------------------------------------------------

Theoretical models describe complex dust processing mechanisms in SNRs. When reverse shocks impact SN ejecta, dust grains 
undergo destructive processing through gas-grain collisions (sputtering; \citealt{Dwek1996}) and grain-grain collisions 
(shattering; \citealt{Jones1994}). These mechanisms reduce total dust mass and alter size distributions, with outcomes 
depending sensitively on initial dust composition and size \citep{Micelotta2018}. Models consistently show small grains 
are more efficiently destroyed by thermal sputtering, while large grains gradually decrease in size. However, predictions 
diverge: some suggest a shift toward smaller grains \citep[e.g.][]{BS2007}, while others indicate complete destruction of small 
grains leading to larger average sizes \citep[e.g.][]{Nozawa2007}. Forward shocks propagating into surrounding ISM similarly 
destroy ambient dust, preferentially affecting small grains \citep{Bocchio2014}.
% ---------------------------------------------------------------------------------------------------------------------

Obtaining observational constraints to test these models remains challenging. While IR and (sub-)millimeter SED modeling 
provides estimates of dust mass, temperature, and composition in SNRs \citep[e.g.][]{Dwek1987, Barlow2010, Gomez2012}, 
these parameters exhibit significant degeneracies. For example, dust mass estimates in Cas A range from $0.075\,\Msun$ 
\citep{Barlow2010} to $0.4{-}0.6\,\Msun$ \citep{DeLooze2017} depending on composition assumptions. Furthermore, temporal 
variations of SN ejected dust mass appear, as evidenced by SN 1987A's reported dust mass increase from ${\sim}10^{-3}\,\Msun$ 
to $0.4{-}0.7\,\Msun$ over two decades \citep{Matsuura2011, Wesson2015, Matsuura2015}. Crucially, SED modeling is generally 
insensitive to grain size distribution--the key parameter altered by SNR shocks.
% ---------------------------------------------------------------------------------------------------------------------

Therefore, studying the wavelength-dependent extinction laws, which provide key constraints on dust properties, is essential 
for fully understanding dust evolution in SNRs. \citet{hz2018} derived a highly consistent near-IR extinction law for 
dust in the Monoceros SNR (G205.5+0.5), Rosette Nebula, and NGC 2264, suggesting thorough mixing between net SN dust and 
ISM dust in older SNRs (e.g., ${\sim}10^5$\,yr for Monoceros; \citealt{Graham1982}). Subsequently, by modeling multi-band 
reddening from optical to near-IR with a simple dust model composed of graphite and silicate, \citet{hz2020} found a 
bifurcated grain size distribution for 22 Galactic SNRs: while graphite grains in most SNRs exhibit an average size $\avea$ 
around 0.008\,{\micron}, silicate grains are generally larger, with $\avea$ up to ${\sim}0.03$\,{\micron}. This could be 
explained by the preferential destruction of small silicate grains by SN shocks. However, the fixed mass ratio assumed 
between graphite and silicate introduced significant uncertainties.
% ---------------------------------------------------------------------------------------------------------------------

This study leverages a newly released three-dimensional (3D) map of the dust extinction parameter $\Rv$ \citep{ZhangXY2025} 
to measure dust properties in 14 Galactic SNRs and their surrounding ISM. Since $\Rv \equiv A_{\rm V} / E(B-V)$ reflects 
the slope of the extinction curve, where higher values indicate a larger fraction of big dust grains \citep{WD01}, variations 
in $\Rv$ across SNRs reveal changes in grain size distribution. By focusing on SNRs with radii between 4 and 15\,pc, as 
constrained by the map's spatial resolution, we target remnants likely in the early Sedov phase, with ages under $10^4$\,yr 
\citep[Eq. 39.9;][]{Draine2011}, assuming a hydrogen density of the surrounding ISM $n_0\,{=}\,1\,{\rm cm^{-3}}$ and a 
kinetic energy $E_{51}\,{=}\,2$ ($E_0\,{=}\,2\,{\times}\,10^{51}$\,erg) based on estimates for Cas A \citep{Young2006,
Orlando2016}. In this phase, both SN and swept-up ISM dust undergo significant shock processing, driving notable changes 
in dust properties. To account for variations in the surrounding ISM, we also measure $\Rv$ for the nearby dust, enabling 
a clearer comparison. The paper is structured as follows: Section \ref{sec:data} details the $\Rv$ map, SNR selection, 
and $\Rv$ estimation for each SNR and its vicinity. Section \ref{sec:result} presents the main results and discussions, 
while Section \ref{sec:summary} summarizes the conclusions and their implications.
% ---------------------------------------------------------------------------------------------------------------------

\section{Data and Method} \label{sec:data}

\subsection{Target selection} \label{subsec:target}

Our primary SNR sample was selected from \citet{Yu2019} and \citet{hz2020}, who determined SNR distances using multi-band 
dust extinctions. By cross-referencing these results with literature values and considering associations between SNRs and 
molecular clouds (MCs), we identified ten SNRs with reliable distance estimates and physical radii $\Rsnr\,{<}\,15$\,pc. 
The radii were calculated using the distances and major angular diameters ($\Tsnr$) reported by \citet{Green2025}. To expand 
our sample, we further searched for additional SNRs meeting these criteria and possessing well-constrained distances. Four 
SNRs were subsequently added: DA 495 (G65.7+1.2), a pulsar wind nebula (PWN) with a distance of 1.5\,kpc derived from {\HI} 
polarization and a flat rotation model \citep{Kothes2004}\footnote{Although \citet{Kothes2008} proposed a smaller distance 
to DA 495 with $\Rsun\,{=}\,7.6$\,kpc, the value of $\Rsun$ adopted by \citet{Kothes2004} is more consistent with recent 
determinations \citep[e.g.][]{Reid2019}.}; G114.3+0.3, assigned to the local arm at 0.7\,kpc based on optical emission 
and {\HI} observations \citep{Yar-Uyaniker2004}; Puppis A (G260.4--3.4), with a kinematic distance of $1.4\,{\pm}\,0.1$\,kpc 
inferred from associations with both {\HI} \citep{Reynoso2017} and CO \citep{Aruga2022}; and Vela (G263.9--3.3), a well-studied 
PWN at $287\,{\pm}\,18$\,pc as determined from pulsar parallax measurements \citep{Dodson2003}; Detailed parameters for 
all 14 selected SNRs are provided in Table \ref{tab:snr}. Our selection criteria, particularly the reliance on extinction-based 
distances which require an association with MCs, naturally introduce a bias of our sample towards core-collapse SNRs. 
We note that the SNR sample from \citet{Green2025} does not follow a single surface brightness--diameter relation, which 
implies significant variations in their initial explosion energies ($E_{51}$) and ambient gas densities ($n_0$). This 
inherent physical inhomogeneity is expected to directly influence the dust properties within the remnants and is a source 
of potential uncertainty in our analysis.
% ---------------------------------------------------------------------------------------------------------------------

\subsection{Three-dimensional \texorpdfstring{$\Rv$}{Rv} map} \label{subsec:Rv_map}

In this study, we utilize the 3D $\Rv$ map constructed by \citet{ZhangXY2025}, which is based on low-resolution BP/RP 
spectra \citep{DeAngeli2023,GSP-phot} from Gaia DR3 \citep{Vallenari2023} for 130 million stars. This 3D $\Rv$ map enables 
the determination of local $\Rv$ values for SNRs and their surrounding ISM. The map employs the HEALPix pixelization 
scheme \citep{Gorski2005} and is divided into 25 distance bins extending to $\sim$5\,kpc from the Sun. We accessed the 
map with $N_{\rm side}\,{=}\,128$, corresponding to a spatial resolution of $27^{\prime}$, via Zenodo, doi: 
\href{https://zenodo.org/records/11394477}{10.5281/zenodo.11394477}.
% ---------------------------------------------------------------------------------------------------------------------

\subsection{Deriving \texorpdfstring{$\Rv$}{Rv} in SNRs and their vicinity} \label{subsec:Rv}

\begin{figure*}[ht!]
  \centering
  \includegraphics[width=8cm]{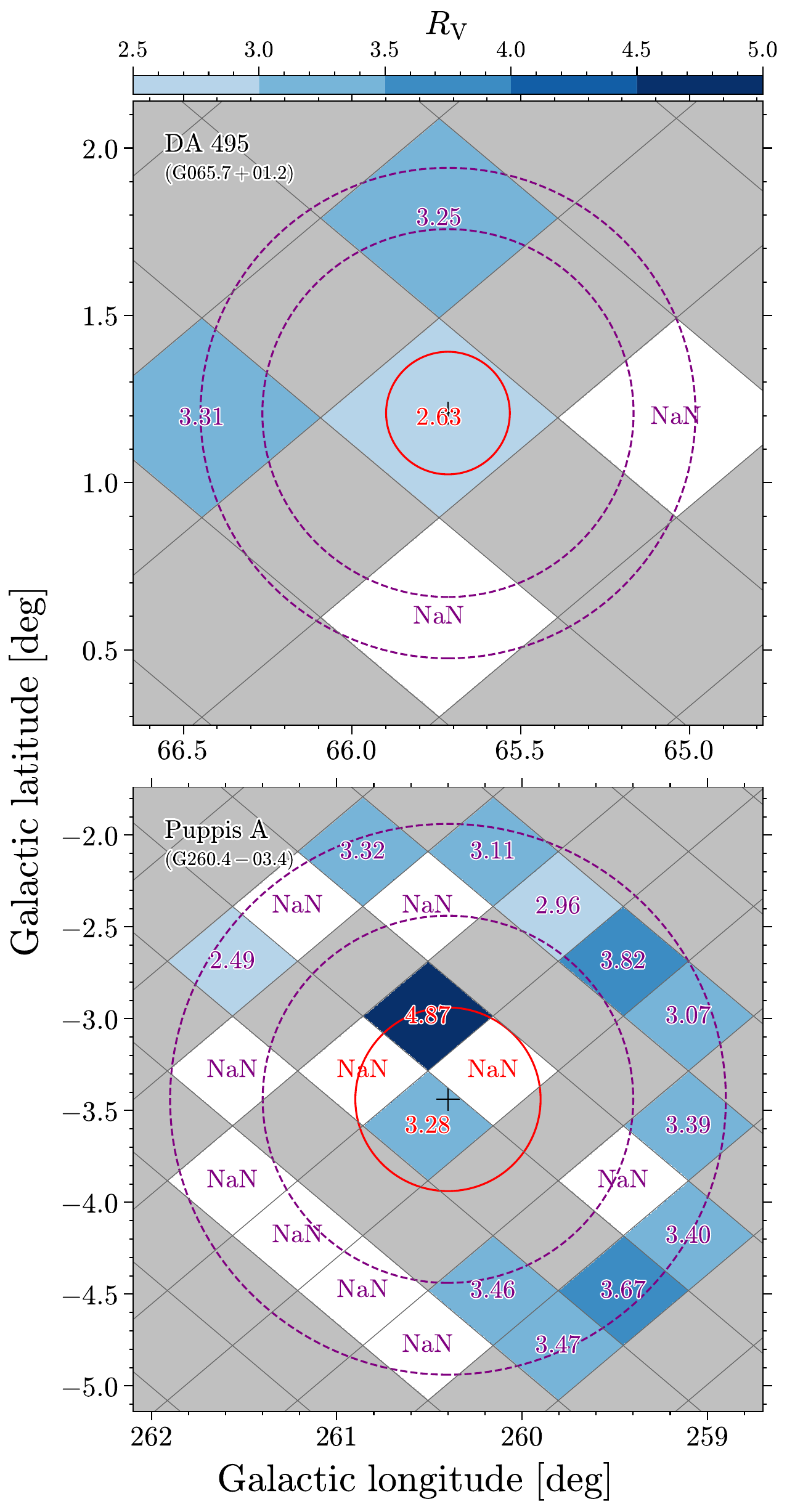}
  \caption{Examples of deriving $\Rv$ in SNRs and their vicinity for SNRs DA 495 (upper) and Puppis A (lower). The red 
  circles and black pluses indicate the SNR angular sizes ($\Tsnr$) and centers. The rhombuses represent the HEALPixels 
  of the used 3D $\Rv$ map ($N_{\rm side}\,{=}\,128$). The HEALPixels with centers in the SNR region or the ISM annulus 
  (dashed purple circles)} are colored and marked by their $\Rv$ values. Empty HEALPixels are marked as `NaN' and colored 
  in white. External HEALPixels are colored in silver.
  \label{fig:1}
\end{figure*}

Given the limited spatial resolution of the $\Rv$ map, we define for each SNR a circular region with the angular diameter 
($\Tsnr$) from \citet{Green2025} to represent the SNR region, and an annular region to represent the surrounding ISM. The 
inner and outer diameters of this annulus are scaled with $\Tsnr$ to ensure the ISM regions are neither overlap with the
SNR circles for small $\Tsnr$ nor too far away for large $\Tsnr$. Specifically, the annular size\footnote{The size relative
to $\Tsnr$ always refers to angular diameters.} is $3{-}4\,\Tsnr$ when $\Tsnr\,{<}\,50^{\prime}$, $2{-}3\,\Tsnr$ when 
$50^{\prime}\,{<}\,\Tsnr\,{<}\,120^{\prime}$, and $1.5{-}2.5\,\Tsnr$ when $\Tsnr\,{>}\,120^{\prime}$. The median dust $\Rv$ 
values are separately derived for these SNR and ISM regions. For each SNR, only the relevant distance slice from the $\Rv$ 
map--corresponding to the SNR's location--is used, as all SNRs in our sample have physical diameters smaller than the map's 
distance bin width. HEALPixels whose centers fall within the SNR circle or ISM annulus are selected to compute $\Rv$ values, 
ensuring no overlap between the pixels used for the SNR and those for the ISM. Figure \ref{fig:1} illustrates the defined 
regions and the corresponding $\Rv$ distributions for DA 495 and Puppis A. 

These examples highlight two key sources of uncertainty driven by the map's resolution. For compact sources like DA 495 
($\Tsnr\,{=}\,22^{\prime}$), which falls within a single pixel, the derived $\Rv$ is subject to potential contamination 
from the surrounding ISM. For more extended sources like Puppis A, which spans four pixels, discretization effects at the 
boundary become apparent: the circular region does not perfectly align with the HEALPix grid, causing some selected pixels 
to extend beyond the remnant while omitting other partially covered regions. These effects--beam dilution for compact sources 
and pixelation artifacts for extended ones--combined with the intrinsic morphological irregularities of the SNRs, introduce 
systematic uncertainties into our $\Rv$ measurements. We applied different SNR and ISM regions to evaluate these impacts
and confirmed that our derived $\Rv$ values are robust and primarily reflect dust properties internal to the remnants (see 
Section \ref{subsec:reliable} for a detailed analysis).
% ---------------------------------------------------------------------------------------------------------------------

Since the $\Rv$ map does not provide uncertainties for individual HEALPixels, we estimate the median $\Rv$ value and its 
statistical uncertainty for each SNR using the 50th, 16th, and 84th percentiles from a 2000-time bootstrap resampling. 
The resulting median values from the SNR circle and ISM annulus are referred to as the SNR $\Rv$ and ISM $\Rv$, respectively, 
in the following discussion. The derived SNR and ISM $\Rv$ values for each target are also listed in Table \ref{tab:snr}.
% ---------------------------------------------------------------------------------------------------------------------

\section{Results and Discussions} \label{sec:result}

\subsection{Grain size evolution traced by \texorpdfstring{$\dRv$}{dRv}} \label{subsec:trend}

\begin{figure*}[ht!]
  \centering
  \includegraphics[width=14cm]{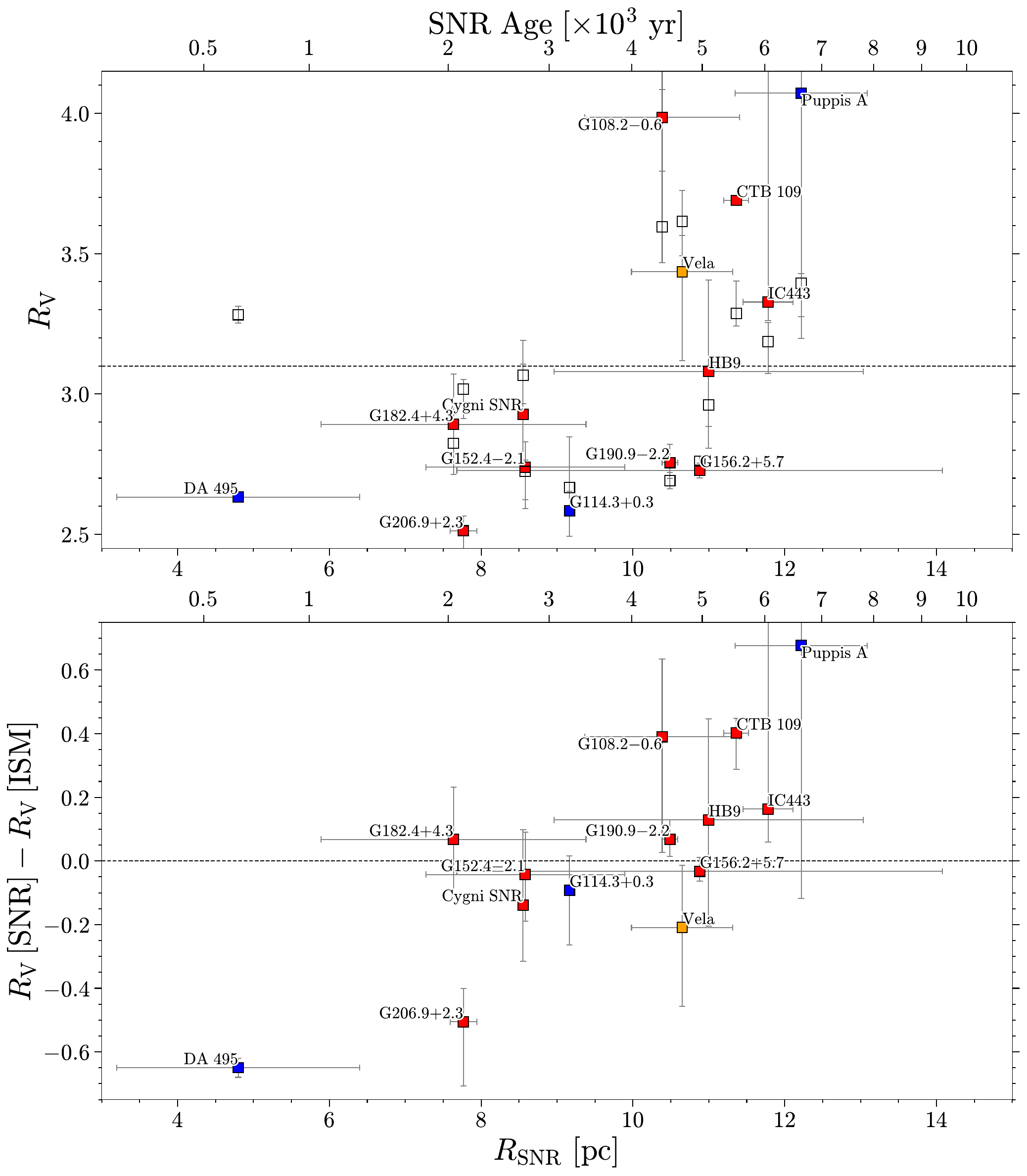}
  \caption{$\Rv$ values as a function of SNR radii $\Rsnr$. The upper panel shows the SNR $\Rv$ (colored squares) and the 
  ISM $\Rv$ (white squares) for 14 SNRs whose widely used names are marked as well. The lower panel shows the differences 
  between the SNR and ISM $\Rv$. The colors represent the different methods of determining SNR distances: Red: extinction; 
  Blue: kinematics; Orange: association with known objects. The SNR ages are also indicated in the top in each panel, 
  estimated by a simple age--diameter relation in Sedov phase (Eq. 39.9; \citealt{Draine2011}). The dashed black lines 
  indicate $\Rv\,{=}\,3.1$ and $\dRv\,{=}\,0$, respectively.}
  \label{fig:2}
\end{figure*}

Our analysis reveals a systematic relationship between dust properties and SNR size. As shown in the upper panel of Figure
\ref{fig:2}, SNR $\Rv$ values generally increase with $\Rsnr$. Small SNRs with radii $\Rsnr\,{\lesssim}\,10$\,pc 
consistently exhibit $\Rv$ values below the Milky Way average of $\Rv\,{=}\,3.1$. In contrast, larger SNRs typically 
exhibit $\Rv$ values exceeding 3.1. However, the $\Rv$ of the surrounding ISM varies across different regions of the Galaxy, 
complicating direct comparisons. For most small SNRs, the surrounding ISM also has $\Rv\,{\lesssim}\,3.1$, with the notable 
exception of DA 495, where the ISM $\Rv$ is approximately $3.3$. To account for these regional variations, we use the 
difference between the SNR and ISM $\Rv$ values, defined as $\dRv = {\rm SNR}\,\Rv - {\rm ISM}\,\Rv$, as a more 
reliable indicator of dust property changes in young Sedov-phase SNRs. We calculate $\dRv$ and its uncertainty using 
bootstrap resampling, rather than simply subtracting median $\Rv$ values, to ensure robust statistical analysis (see 
Table~\ref{tab:snr}).
% ---------------------------------------------------------------------------------------------------------------------

As shown in the lower panel of Figure \ref{fig:2}, $\dRv$ reveals a more consistent and systematic dependence on $\Rsnr$ 
compared to the absolute SNR $\Rv$. For instance, DA 495 shares similar SNR $\Rv$ values with G114.3+0.3 and G206.9+2.3 
but has a significantly higher ISM $\Rv$ ($3.28\,{\pm}\,0.03$), resulting in the smallest $\dRv$ among these SNRs. These 
three remnants exhibit a monotonic increase in $\dRv$ with increasing $\Rsnr$. Across the full sample, $\dRv$ rises from 
${-}0.65$ at $\Rsnr \sim 5$\,pc to a maximum of 0.68 at $\Rsnr\,{\sim}\,12$\,pc, with a transition point around $\Rsnr\,{\sim}\,10{-}11$\,pc 
where SNR $\Rv$ begins to surpass ISM $\Rv$. Extrapolating this trend to earlier SNR evolutionary stages aligns with 
observations of extragalactic SNe, which typically show low $\Rv$ values of $1.5{-}2.5$ \citep[e.g., SNe 2001da, 2006cm, 
2007fb, 2007on, 2010ev in][]{Phillips2013}. Moreover, \citet{Rino-Silvestre2025} reported higher $\Rv \sim 3.1$ values 
in the vicinity of these SNe using optical photopolarimetry, suggesting negative $\dRv$ as well, though these measurements 
may be affected by ISM contamination \citep{Phillips2013} or color biases \citep{Scolnic2014}.
% ---------------------------------------------------------------------------------------------------------------------

The majority of our sample (11 of 14 SNRs) follow the trend that SNRs with $\Rv\,{>}\,3.1$ also have $\dRv\,{>}\,0$, and 
vice versa. G190.9--2.2 has an SNR $\Rv$ below 3.1 yet a slightly positive $\dRv\,{=}\,0.07_{-0.05}^{+0.06}$, a behavior 
consistent with its position along the broader $\dRv-\Rsnr$ trend. The two more notable exceptions are G182.4+4.3, which 
displays a positive $\dRv$ despite its small radius ($\Rsnr\,{=}\,7.6\,{\pm}\,1.7$\,pc), and Vela, which shows a negative 
$\dRv\,{=}\,{-}0.21_{-0.25}^{+0.20}$ at a relatively large radius ($\Rsnr\,{=}\,10.6\pm 0.7$\,pc). In addition to measurement 
uncertainties in distance and $\Rv$, variations in initial conditions (e.g., $E_{51}$ and $n_0$) can influence SNR expansion 
and dust evolution, potentially resulting in different $\dRv$ values among SNRs of similar $\Rsnr$. The Spearman correlation 
coefficient ($\rs$) between $\dRv$ and $\Rsnr$ is 0.75 for the full sample. When incorporating measurement uncertainties 
through Monte Carlo resampling, this decreases to $0.62\,{\pm}\,0.14$, indicating that current uncertainties 
moderately impact--but do not invalidate--the observed trend, which remains statistically robust.
% ---------------------------------------------------------------------------------------------------------------------

\subsection{Grain size redistribution under shock processing} \label{subsec:grain_size}

The observed increase in both SNR $\Rv$ and $\dRv$ with $\Rsnr$ suggests a shift in the dust grain size distribution 
toward larger grains as SNRs expand, in contrast to cold, dense environments where $\Rv$ increases due to dust growth 
processes \citep{Li2024a, Li2024b}. This trend arises from distinct dust processing mechanisms that dominate at different 
evolutionary stages. In compact remnants ($\Rsnr\,{\lesssim}\,10$\,pc), the gradual increase of $\dRv$, while remaining 
below zero, reflects the impact of reverse shocks on newly formed SN dust. Reverse shocks efficiently destroy the small 
grains through thermal sputtering \citep{Dwek1996}, increasing the average grain size slightly but leaving the system 
deficient in large grains (${>}\,0.1$\,{\micron}) compared to the ISM. Since $\Rv$ depends critically on the absolute 
abundance of large grains \citep{WD01}, the processed dust maintains $\Rv\,{<}\,3.1$ while the surrounding ISM retains 
its typical $\Rv\,{\gtrsim}\,3.1$, resulting in negative $\dRv$ as observed in DA 495 ($\dRv\,{=}\,{-}0.65\,{\pm}\,0.03$).
% ---------------------------------------------------------------------------------------------------------------------

As remnants expand beyond $\Rsnr\,{\sim}\,10$\,pc, two transformative processes occur: 1) swept-up ISM dust 
begins to dominate the total dust mass ($\propto \Rsnr^3$), and 2) forward shocks selectively process this ambient dust. 
Unlike SN ejecta dust, the ISM contains abundant large grains. Forward shocks preferentially destroy small ISM grains 
(${\lesssim}\,0.01$\,{\micron}) through sputtering and shattering \citep{Bocchio2014}, further enhancing the relative 
abundance of large grains. This grain size redistribution elevates the SNR $\Rv$ above both its initial value and the 
surrounding ISM value, producing positive $\dRv$ as seen in Puppis A ($\dRv\,{=}\,0.68_{-0.80}^{+0.83}$).
% ---------------------------------------------------------------------------------------------------------------------

Our findings are consistent with the theoretical framework developed by \citet{SC2015} and \citet{Nozawa2007,Nozawa2010}. 
In particular, \citet{Nozawa2007} predicted that dust processing in SNRs would establish a critical size threshold 
($a_{\rm crit}\,{\lesssim}\,0.05$\,{\micron} for the dust species considered therein), below which grains are completely 
destroyed. This mechanism naturally produces the observed shift toward larger grains and increasing $\Rv$ with remnant age. 
In contrast, models by \citet{BS2007} that allow survival of very small grains through a flattened size distribution are 
inconsistent with our data, as they would predict lower $\Rv$ values than observed. The observed inflection point at 
$\Rsnr\,{\sim}\,10$\,pc corresponds to the transition from ejecta-dust to ISM-dust dominance, with its exact position 
modulated by initial conditions ($E_{51}$, $n_0$) as evidenced by the outlier G182.4+4.3, which may have experienced 
enhanced processing due to higher ambient density. The presence of a PWN can also influence dust evolution by injecting 
additional energy that ionizes the ejecta gas and sublimates dust seeds, potentially delaying the dust formation and 
reducing the average grain size \citep{Omand2019}. However, this effect is thought to be significant only for pulsars 
with very rapid initial rotation periods ($P_{\rm init}\,{<}\,10$\,ms). Of the two PWNe in our sample, the pulsar in 
DA 495 remains unconfirmed \citep{Lorimer1998}, while the Vela pulsar's estimated $P_{\rm init}\,{\approx}\,30$\,ms, 
given the present $P\,{=}\,89$\,ms and a spin-down rate of $1.24\,{\times}\,10^{-13}\,{\rm s\,s^{-1}}$ \citep{Grondin2013,
Lange2025}, is well above this threshold. As a result, while PWN effects cannot be entirely ruled out, they are less 
likely to be a dominant driver of the dust evolution trends observed across our sample.
% ---------------------------------------------------------------------------------------------------------------------

For a more quantitative analysis, we assume $n_0\,{=}\,1\,{\rm cm^{-3}}$ and a gas-to-dust ratio of 100 to estimate the 
mass of swept-up ISM dust ($\MswDust$) prior to destruction. For DA 495, $\MswDust\,{\sim}\,0.11\,\Msun$, while for Puppis A, 
$\MswDust$ increases to $1.88\,\Msun$. By comparison, the mass of dust formed in the ejecta ($\MejDust$) of SN 1987A is 
estimated to be $0.4{-}0.7\,\Msun$ \citep{Matsuura2011, Matsuura2015} about 20 years after explosion, before significant 
reverse shock interaction. Similarly, \citet{DeLooze2017} found $0.4{-}0.6\,\Msun$ of cold dust in the unshocked region 
of Cas A. This substantial SN dust mass is comparable to the swept-up ISM dust at $\Rsnr \sim 8{-}9$\,pc. Additionally, 
the very small $\Rv$ values observed in extragalactic SNe suggest that SN ejecta dust is initially dominated by small 
grains. Assuming comparable destruction efficiencies of forward and reverse shocks on SN and ISM dust, the initial increase 
in $\Rv$ for small SNRs at the onset of the Sedov phase (e.g., DA 495, $\Rsnr\,{=}\,4.8$\,pc) is primarily attributed to the 
destruction of small SN dust grains by the reverse shock, though the $\Rv$ remains lower than that of the surrounding ISM 
($\dRv\,{<}\,0$). As more ISM dust--where large grains are more likely to survive the forward shock--is swept up, $\dRv$ 
continues to increase for $\Rsnr\,{\sim}\,8{-}10$\,pc. For larger SNRs such as Puppis A ($\Rsnr\,{=}\,12.2$\,pc), where 
the swept ISM dust mass greatly exceeds that of the SN dust and dominates the dust properties, the SNR $\Rv$ increases 
further and surpasses the ISM value ($\dRv\,{>}\,0$). The positive $\dRv\,{=}\,0.68_{-0.80}^{+0.83}$ for Puppis A is 
consistent with the findings of \citet{Arendt2010}, who used {\it Spitzer}/IRS observations to show that small grains 
in compact knots of Puppis A have been largely destroyed. Of course, varying initial conditions among SNRs can lead to 
diverse evolutionary paths. For instance, the $\dRv$ values of G182.4+4.3 and G108.2--0.6 are noticeably above the overall 
$\dRv$--$\Rsnr$ trend, potentially due to a lower $\MejDust$ or higher $n_0$. Nevertheless, the moderate $\rs$ between 
$\dRv$ and $\Rsnr$ may suggest a general scenario for dust evolution in SNRs.
% ---------------------------------------------------------------------------------------------------------------------

The ages of SNRs ($\tsnr$) in the Sedov phase can be estimated using Eq.\,39.9 of \citet{Draine2011}. Assuming 
$E_{51}\,{=}\,2$ and $n_0\,{=}\,1\,{\rm cm^{-3}}$, we indicate $\tsnr$ at the top of each panel in Figure \ref{fig:2} 
for various $\Rsnr$ values. Although these simple estimates may differ from literature values (e.g., $\tsnr\,{\sim}\,6\,000$\,yr 
for IC443 is smaller than previous estimates; see \citealt{Olbert2001, Lee2008}), the interaction timescale between SN 
shocks and dust is generally less than $10^4$\,yr for our SNR sample. Therefore, it is important to consider the rates 
and timescales of grain destruction for different sizes when interpreting dust evolution. According to the models of 
\citet{Nozawa2007}, the collision time of the reverse shock with dust is about $10^3$\,yr for a $20\,\Msun$ progenitor 
and $n_0\,{=}\,1\,{\rm cm^{-3}}$, but this can decrease to several hundred years if the progenitor has experienced 
significant hydrogen envelope loss (such as Cas A; \citealt{Lee2014,Orlando2016}). This nominal $10^3$\,yr collision 
timescale implies that the reverse shock has already swept through and processed the ejecta dust in all but the youngest 
remnant in our sample, DA 495. On the other hand, the lifetimes of 0.01\,{\micron} grains are between ${\sim}10^4$ (for 
silicate) and ${\sim}\,10^5$\,yr (for carbon), and exceed $10^5$\,yr for 0.05\,{\micron} grains of all major dust species 
(carbon, silicate, and Fe). While these lifetimes exceed the ages of our SNRs, the dust erosion by thermal sputtering 
is a continuous process. Simulations by \citet{Nozawa2007} show that this process is highly size-dependent: erosion of 
0.01\,{\micron} grains begins at just 3\,000\,yr, with this timescale becoming even shorter for smaller grains, while 
grains larger than 0.1\,{\micron} remain almost unaffected over much longer timescales. Moreover, the erosion rate for 
small grains is faster as well, ensuring they are substantially processed within the typical ${\sim}10^4$\,yr ages of 
our SNRs. These theoretical predictions are consistent with our observations: in the early Sedov phase, the reverse 
shock is efficient at destroying the smallest SN dust grains, while a significant fraction of larger grain population 
can survive for over $10^4$\,yr. This net shift toward larger average grain sizes can explain the gradual increase of 
$\dRv$ in the regime where $\Rsnr\,{\lesssim}\,10$\,pc. Since time-resolved grain size distributions under shock processing 
are not currently available, we assume a similar destruction timescale for ISM dust under forward shock to explain the 
continuous increase of $\dRv$ for $\Rsnr\,{\gtrsim}\,10$\,pc.
% ---------------------------------------------------------------------------------------------------------------------

\subsection{Reliability and limitations} \label{subsec:reliable}

\begin{figure*}[ht!]
  \centering
  \includegraphics[width=14cm]{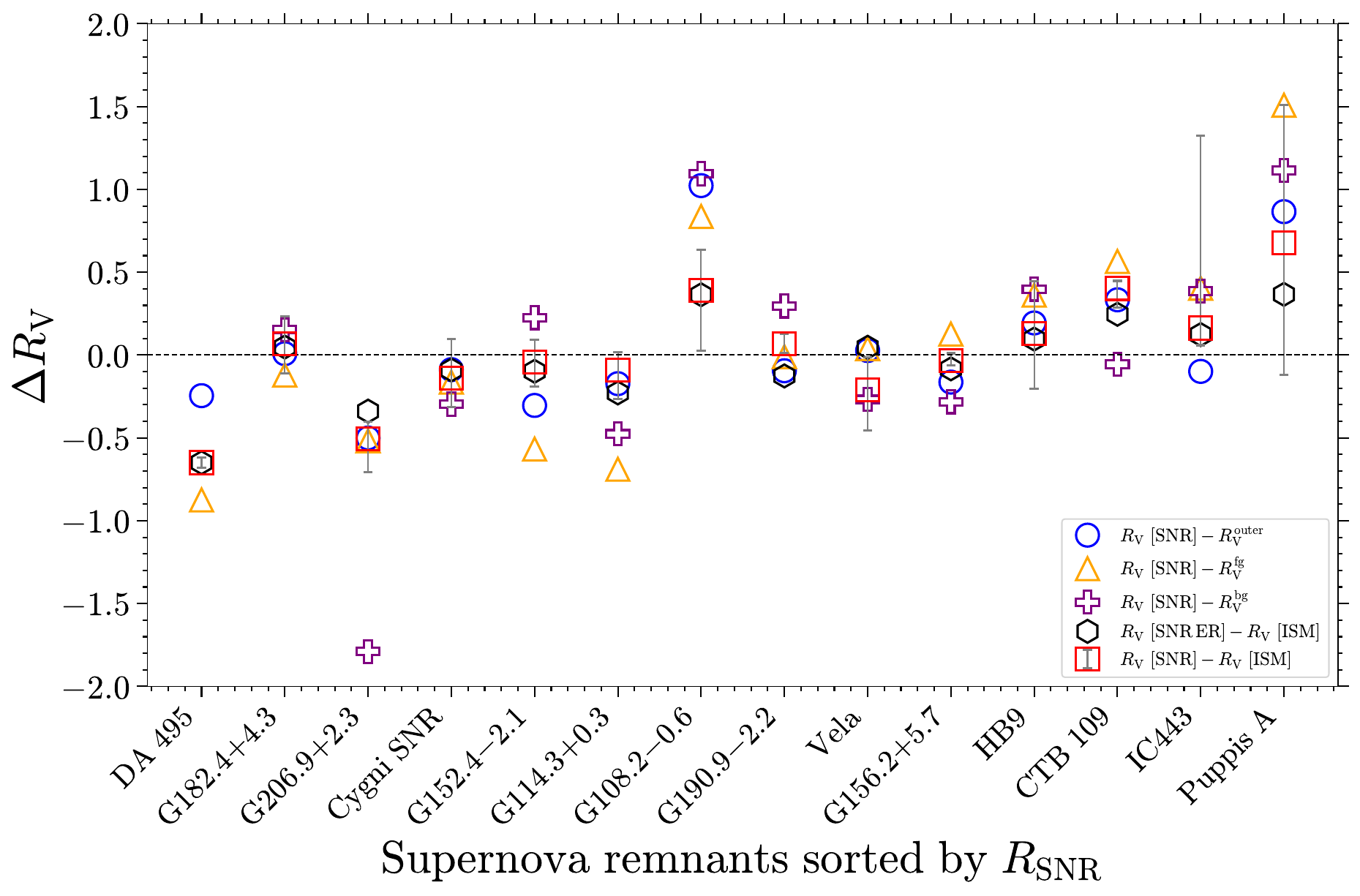}
  \caption{Comparison of $\Delta \Rv={\rm SNR}\,\Rv-{\rm ISM}\,\Rv$ with ISM $\Rv$ derived from different regions relative
  to the SNRs. Red square: an annular region with dynamic ranges (see Section \ref{subsec:Rv}) at the SNR distance; Blue 
  circle: a bigger annulus ($4{-}5\,\Tsnr$) at the SNR distance; Orange triangle: a ring ($\Tsnr$) in front of the SNR (in 
  a foreground distance bin in the 3D $\Rv$ map); Purple plus: a ring ($\Tsnr$) behind the SNR (in a background distance 
  bin in the 3D $\Rv$ map). Black hexagons: an annular region of $3{-}4\,\Tsnr$ for ISM with an expanded region (ER) for 
  SNR ($1.6\,\Tsnr$).} 
  \label{fig:3}
\end{figure*}

The reliability of our $\Rv$ measurements warrants careful consideration. Since all SNR diameters in our sample are smaller 
than the distance bin sizes (${\sim}100{-}200$\,pc) of the \citet{ZhangXY2025} $\Rv$ map, we derived each SNR's $\Rv$ from 
a single map slice. This approach raises potential concerns about contamination from unrelated ISM along the line of sight. 
To evaluate this, we calculated median $\Rv$ values in adjacent distance bins along the same sightline--denoted $\Rv^{\rm fg}$ 
(foreground distance bin) and $\Rv^{\rm bg}$ (background distance bin)--within the same projected region ($\Tsnr$). While 
differences between SNR $\Rv$ and $\Rv^{\rm fg}$/$\Rv^{\rm bg}$ vary across remnants (orange triangles and purple pluses 
in Figure \ref{fig:3}), they generally follow the same trend as $\dRv$ (red squares). This consistency confirms that the 
SNR $\Rv$ measurements predominantly reflect local dust properties rather than broader sightline effects. To further test 
the robustness of our results against aperture effects and pixelation artifacts, we recalculated $\dRv$ using an expanded 
SNR region ($1.6\,\Tsnr$) with an ISM annular region extending from $3\,\Tsnr$ to $4\,\Tsnr$. The resulting trend (black 
hexagons in Figure \ref{fig:3}) remains consistent with our primary findings. Taken together, these tests demonstrate that 
the observed trends are robust and primarily driven by dust processing within the SNRs. Nevertheless, some degree of 
contamination from ambient ISM within the same distance bin is unavoidable. This effect likely dilutes the true contrast 
between the SNR and its environment, meaning our reported $\dRv$ values may be conservative estimates of the actual effect.
% ---------------------------------------------------------------------------------------------------------------------

To evaluate the impact of spatial variations in ISM dust properties, we calculated the median $\Rv$ from a wider ISM annulus 
($4{-}5\,\Tsnr$), referred to as $\Rv^{\rm outer}$. For most SNRs, the difference between SNR $\Rv$ and $\Rv^{\rm outer}$ 
(blue circles in Figure \ref{fig:3}) is consistent with $\dRv$ within uncertainties, with the largest deviation found for 
G108.2--0.6. Overall, while variable ISM dust properties may increase the scatter and uncertainties, they do not obscure 
the observed $\dRv{-}\Rsnr$ trend. 
% ---------------------------------------------------------------------------------------------------------------------

The intrinsic diversity in SNR initial conditions ($E_{51}$, $n_0$) is a fundamental source of physical scatter. This 
diversity impacts not only the dust evolution itself but also the validity of applying a single age-radius relation 
across the sample. Consequently, the scatter observed in the $\dRv-\Rsnr$ trend is an expected feature of the underlying 
physics. We anticipate that future work incorporating accurate measurements of these parameters could reveal a tighter, 
more fundamental relationship, likely through the application of appropriate scaling factors.
% ---------------------------------------------------------------------------------------------------------------------

In addition to the physical scatter introduced by initial conditions and the finite resolution of the $\Rv$ map, a third 
major limitation is the uncertainty in SNR distances. Although we have selected nearby SNRs with relatively reliable 
distance measurements, the mean distance uncertainty remains around 15\%, and exceeds 20\% for G156.2+5.7, G182.4+4.3, 
and DA 495. These uncertainties can affect the interpretation of dust evolution within SNRs. Achieving more precise 
distance determinations for SNRs remains an important but challenging goal for future work. Furthermore, a more detailed 
analysis of dust evolution, incorporating variations in SN initial conditions and dust composition, will be conducted in 
subsequent studies.
% ---------------------------------------------------------------------------------------------------------------------

\section{Conclusions} \label{sec:summary}

By leveraging the newly released 3D $\Rv$ map, we have measured local $\Rv$ values for dust in 14 early Sedov-phase SNRs 
in the Milky Way and their surrounding ISM, effectively minimizing foreground and background contamination. We find a 
moderately strong positive correlation between the difference in SNR and ISM $\Rv$ ($\dRv$) and SNR radius ($\Rsnr$), 
with a Spearman coefficient of $\rs\,{=}\,0.62\,{\pm}\,0.14$. This trend provides direct observational evidence 
for a redistribution of dust grain sizes toward larger grains under SN shock processing--consistent with theoretical 
predictions by \citet{SC2015}, \citet{Nozawa2007,Nozawa2010}, and \citet{Bocchio2014}. Our results indicate that, for 
small SNRs such as DA 495 with $\Rsnr\,{=}\,4.8\,{\pm}\,1.6$\,pc, the increase of $\dRv$ is primarily driven by the 
destruction of small SN dust grains by the reverse shock, while for larger SNRs like Puppis A ($\Rsnr\,{=}\,12.2\,{\pm}\,0.9$\,pc), 
the dominance of swept-up ISM dust--where larger grains preferentially survive the forward shock--leads to $\dRv\,{>}\,0$. 
These findings provide essential observational constraints on the dust processing and grain size evolution in SNRs.
% ---------------------------------------------------------------------------------------------------------------------

Our results offer significant insights into the rapid formation and evolution of dust in the early Universe, where SNe 
are thought to be the primary dust sources. By demonstrating how shock processing alters dust grain sizes in local SNRs, 
this study provides a critical link to understanding dust survival and enrichment during cosmic dawn. The observed shift 
toward larger grains in expanding SNRs suggests that dust injected into the early ISM may be more resilient than previously 
thought, influencing the spectral properties and detectability of high-redshift galaxies. Looking forward, higher-resolution 
$\Rv$ maps and multi-wavelength observations of SNRs could further constrain dust composition and size distributions, 
enabling more precise comparisons with theoretical models. Additionally, extending similar analyses to extragalactic SNRs 
with upcoming facilities, such as the James Webb Space Telescope (JWST) and Chinese Space Station Telescope (CSST), could 
directly probe dust evolution in environments closer to those of the early universe, deepening our understanding of cosmic 
dust cycles.
% ---------------------------------------------------------------------------------------------------------------------

\begin{acknowledgments}

We thank the anonymous referee for a thorough review and valuable feedback, which helped to reinforce our results and 
better articulate the limitations of the study. We thank Professor Biwei Jiang for very helpful suggestions and stimulating 
comments. This work is partially supported by the National Natural Science Foundation of China 12173034, 12322304, and 
12403026. H.Z. acknowledges financial support by the Chilean Government-ESO Joint Committee (Comit\'{e} Mixto ESO-Chile, 
No. annlang23003-es-cl). B.Q.C. acknowledges the National Natural Science Foundation of Yunnan Province 202301AV070002 
and the Xingdian talent support program of Yunnan Province. We acknowledge the science research grants from the China 
Manned Space Project with No. CMS-CSST-2025-A11.
% ---------------------------------------------------------------------------------------------------------------------

\end{acknowledgments}

% \begin{contribution}

% \end{contribution}

\facilities{Gaia}

% \software{
%           }

% \appendix

\bibliography{sample701}{}
\bibliographystyle{aasjournalv7}

\begin{deluxetable*}{ccccccccr}
% \digitalasset
\tablewidth{0pt}
\tablecaption{Detailed information and derived $\Rv$ values for the 14 selected SNRs. \label{tab:snr}}
\tablehead{
  \colhead{SNR} & \colhead{Other name} & \colhead{$\Tsnr$\tablenotemark{a}} & \colhead{Distance (kpc)} & 
  \colhead{Ref.\tablenotemark{b}} & \colhead{$\Rsnr$ (pc)\tablenotemark{c}} & 
  \colhead{SNR $\Rv$\tablenotemark{d}} & \colhead{ISM $\Rv$\tablenotemark{e}} & \colhead{$\dRv$\tablenotemark{f}}
}
\startdata
G65.7+1.2   & DA 495    & 22$^\prime$  & $1.5\pm0.5$  & 1  & $4.8 \pm 1.6$ &$2.63                $ & $3.28_{-0.03}^{+0.03}$ & $-0.65_{-0.03}^{+0.03}$ \\
G78.2+2.1   & Cygni SNR	& 60$^\prime$ & $0.98$        & 2  & $8.6$         &$2.93_{-0.19}^{+0.26}$ & $3.07_{-0.10}^{+0.04}$ & $-0.14_{-0.18}^{+0.24}$ \\
G108.2--0.6 & 	        & 70$^\prime$ & $1.02\pm0.01$ & 2  & $10.4\pm 0.1$ &$3.99_{-0.19}^{+0.19}$ & $3.59_{-0.13}^{+0.49}$ & $ 0.39_{-0.36}^{+0.24}$ \\
G109.1--1.0 & CTB 109	& 28$^\prime$ & $2.79\pm0.04$ & 2  & $11.4\pm 0.2$ &$3.69                $ & $3.29_{-0.05}^{+0.11}$ & $ 0.40_{-0.11}^{+0.05}$ \\
G114.3+0.3  & 	        & 90$^\prime$ & $0.7$         & 4  & $9.2$         &$2.58_{-0.09}^{+0.07}$ & $2.67_{-0.09}^{+0.18}$ & $-0.09_{-0.17}^{+0.11}$ \\
G152.4--2.1 & 	      & 100$^\prime$ & $0.59\pm0.09$& 2, 3 & $8.6 \pm 1.3$ &$2.74_{-0.15}^{+0.03}$ & $2.72_{-0.10}^{+0.11}$ & $-0.04_{-0.15}^{+0.13}$ \\
G156.2+5.7  & 	        & 110$^\prime$ & $0.68\pm0.20$& 2  & $10.9\pm 3.2$ &$2.73_{-0.03}^{+0.03}$ & $2.76_{-0.02}^{+0.02}$ & $-0.03_{-0.03}^{+0.04}$ \\
G160.9+2.6  & HB9	  & 140$^\prime$ & $0.54\pm0.10$& 2, 3 & $11.0\pm 2.0$ &$3.08_{-0.27}^{+0.33}$ & $2.96_{-0.08}^{+0.13}$ & $ 0.13_{-0.33}^{+0.32}$ \\
G182.4+4.3  & 	      & 50$^\prime$ & $1.05\pm0.24$ & 2, 3 & $7.6 \pm 1.7$ &$2.89_{-0.18}^{+0.18}$ & $2.82_{-0.02}^{+0.10}$ & $ 0.07_{-0.18}^{+0.16}$ \\
G189.1+3.0  & IC443   & 45$^\prime$ & $1.80\pm0.05$ & 2, 3 & $11.8\pm 0.3$ &$3.33_{-0.07}^{+1.20}$ & $3.19_{-0.11}^{+0.07}$ & $ 0.16_{-0.10}^{+1.16}$ \\
G190.9--2.2 & 	      & 70$^\prime$ & $1.03\pm0.01$ & 2, 3 & $10.5\pm 0.1$ &$2.76_{-0.06}^{+0.07}$ & $2.69_{-0.03}^{+0.03}$ & $ 0.07_{-0.05}^{+0.06}$ \\
G206.9+2.3  & PKS 0646+06 & 60$^\prime$ & $0.89\pm0.02$ & 2& $7.8 \pm 0.2$ &$2.51_{-0.20}^{+0.05}$ & $3.02_{-0.10}^{+0.03}$ & $-0.51_{-0.20}^{+0.10}$ \\
G260.4--3.4 & Puppis A  & 60$^\prime$ & $1.4\pm0.1$   & 5  & $12.2\pm 0.9$ &$4.07_{-0.80}^{+0.80}$ & $3.39_{-0.20}^{+0.03}$ & $ 0.68_{-0.80}^{+0.83}$ \\
G263.9--3.3 & Vela   & 255$^\prime$ & $0.287\pm0.018$ & 6  & $10.6\pm 0.7$ &$3.44_{-0.32}^{+0.13}$ & $3.61_{-0.12}^{+0.11}$ & $-0.21_{-0.25}^{+0.20}$ \\
\enddata
\tablenotetext{a}{SNR major angular diameter from \citet{Green2025}.}
\tablenotetext{b}{Distance references: (1) \citet{Kothes2004}, (2) \citet{hz2020}, (3) \citet{Yu2019}, (4) \citet{Yar-Uyaniker2004},
                  (5) \citet{Aruga2022}, (6) \citet{Dodson2003}.}
\tablenotetext{c}{SNR radius calculated by distance and $\Tsnr$.}
\tablenotetext{d}{Median $\Rv$ and uncertainties derived in SNRs.}
\tablenotetext{e}{Median $\Rv$ and uncertainties derived in surrounding ISM (see ranges in Section \ref{subsec:Rv}).}
\tablenotetext{f}{Differences between SNR and ISM $\Rv$ values.}
\end{deluxetable*}
% ---------------------------------------------------------------------------------------------------------------------

% ---------------------------------------------------------------------------------------------------------------------

% ---------------------------------------------------------------------------------------------------------------------

% ---------------------------------------------------------------------------------------------------------------------

\clearpage
\end{CJK*}
\end{document}